\RequirePackage{fix-cm}
\documentclass[smallextended]{svjour3}       
\smartqed  
\usepackage{graphicx}
\usepackage{natbib}

\begin{document}

\title{Altered connectedness of the brain chronnectome during the progression to Alzheimer’s disease 
}


\author{Maryam Ghanbari         \and
        Zhen Zhou \and 
        Li-Ming Hsu \and Ying Han \and Yu Sun \and Pew-Thian Yap \and Han Zhang* \and Dinggang Shen*
}


\institute{Maryam Ghanbari \at
              Department of Radiology and BRIC, University of North Carolina at Chapel Hill, Chapel Hill, NC, USA
           \and
          Zhen Zhou \at
              Department of Radiology and BRIC, University of North Carolina at Chapel Hill, Chapel Hill, NC, USA 
           \and
           Li-Ming Hsu \at
              Department of Radiology and BRIC, University of North Carolina at Chapel Hill, Chapel Hill, NC, USA 
           \and
            Ying Han \at
              Department of Neurology, Xuanwu Hospital of Capital Medical University, Beijing, China, 100053 \\
              {Center of Alzheimer’s Disease, Beijing Institute for Brain Disorders, Beijing, China, 100053}\\
{Beijing Institute of Geriatrics, Beijing, China, 100053}\\
{National Clinical Research Center for Geriatric Disorders, Beijing, China, 100053}
           \and
           Yu Sun \at
              Department of Neurology, Xuanwu Hospital of Capital Medical University, Beijing, China, 100053 
           \and
          Pew-Thian Yap \at
              Department of Radiology and BRIC, University of North Carolina at Chapel Hill, Chapel Hill, NC, USA
           \and
           Han Zhang*   \at
              Department of Radiology and BRIC, University of North Carolina at Chapel Hill, Chapel Hill, NC, USA \\
              \email{hanzhang@med.unc.edu}           
           \and
           Dinggang Shen*  \at
              Department of Radiology and BRIC, University of North Carolina at Chapel Hill, Chapel Hill, NC, USA \\
              {Department of Brain and Cognitive Engineering, Korea University, Seoul, Republic of Korea }
              \email{idea.uncch@gmail.com}             
                        }

\date{Received: date / Accepted: date}

\maketitle

 

 \section*{Information Sharing Statement}
The source code of the presented method is freely available for use from https://github.com/mghanba/Maryam\_Ghanbari\_Repository/tree/master.  

 

 \section*{Declarations}
{\bf Funding}: M.G. was supported  by the National Institutes of Health grants (EB022880, AG041721 and AG042599). Z.Z., L.-M.H., P.-T.Y. and D.S.  were supported  by the National Institutes of Health grant (EB022880).
Y.H. and Y.S. were supported by National Natural Science Foundation of China (Grants 61633018, 31371007).
H.Z. was supported  by the National Institutes of Health grants (EB022880, AG041721, AG049371, and AG042599).\\
\noindent {\bf Conflicts of interest/Competing interests:} The authors declare that they have no conflict of interest.\\
\noindent {\bf Ethics approval:} The experiments and data collection were approved by the local ethics committees, as mentioned in ADNI data sharing website http://ad ni.loni.usc.edu. For the Xuanwu hospital's data, ethical approval has been obtained from the medical research ethics committee and institutional review board of XuanWu Hospital, Capital Medical University (approval number: [2014]011).\\
\noindent {\bf Consent to participate:} Data used from ADNI is publicly available, so this is not applicable. For the Xuanwu hospital's data, all participation is based on written informed consent and the participants will be able to withdraw from the study at any time.\\
\noindent {\bf Consent for publication:} The publisher
has the permission from the authors to publish the paper\\
\noindent {\bf Availability of data and material:} The time series data from all the subjects as well as the calculated redundancy measurements that support our claims are publicly available at  https://github.com/mghanba/Maryam\_Ghan bari\_Repository/tree/master, upon the manuscript is entering review process.\\
\noindent {\bf Code availability:} The software we used to calculate connectedness and bi-connectedness is SAGE 8.6 (https://www.sagemath.org). The core function for calculating dynamic redundancy statuses and their transitions are publicly available at 
https://github.com/mghanba/Maryam\_Ghanbari\_Repository/tree /master, upon the manuscript is entering review process.\\
\noindent {\bf Authors' contributions:} H.Z. and D.S. designed and conceptualized the study and revised the manuscript.   M.G. drafted and edited the manuscript, analyzed data, interpreted results. H.Z. played a major role in the interpretation of the  results and revision of  the manuscript. L.-M.H. analyzed the data and revised the manuscript. Z.Z. and P.-T.Y. analyzed data and revised the manuscript. Y.H. and Y.S. collected and analyzed part of the data and revised the manuscript. All authors read and approved the final manuscript.

%
\newpage
\begin{abstract}
Graph theory has been extensively used to investigate brain network topology and its changes in disease cohorts. However, many graph theoretic analysis-based brain network studies focused on the shortest paths or, more generally, cost-efficiency. In this work, we use two new concepts, connectedness and 2-connectedness, to measure different global properties compared to the previously widely adopted ones. We apply them to unravel interesting characteristics in the brain, such as redundancy design and further conduct a time-varying brain functional network analysis for characterizing the progression of Alzheimer’s disease (AD). Specifically, we define different (2-)connectedness states and evaluate their dynamics in AD and its preclinical stage, mild cognitive impairment (MCI), compared to the normal controls (NC). Results indicate that, compared to MCI and NC, brain networks of AD tend to be more frequently connected at a sparse level. For MCI, we found that their brains are more likely to be 2-connected in the minimal connected  state as well 
indicating increasing redundancy in brain connectivity. Such a redundant design could ensure maintained connectedness of the MCI’s brain network in the case that pathological attacks break down any link or silenced any node, making it possible to preserve cognitive abilities. Our study suggests that the redundancy in the brain functional chronnectome could be altered in the preclinical stage of AD. The findings can be successfully replicated in a re-test study and with an independent MCI dataset. Characterizing redundancy design in the brain chronnectome using (2-)connectedness analysis provides a unique viewpoint for understanding disease affected brain networks.
\end{abstract}

\section{Introduction}
The human brain can be modeled as a complex network or graph based on various connectivity metrics, such as functional connectivity (FC, denoting edge weights) that is interpreted as interactions or coordination among different brain regions (denoting nodes) (\citealt{sporns2013structure}). 
There have been various means to construct brain functional connectome, such as Pearson’s correlation of the resting-state fMRI (rs-fMRI) signals, and it has been recently extended to time-varying (non-stationary or dynamic) connectome, or chronnectome (\citealt{Chang,Musso,Sakoglu,Cribben,Yuan,Hutchison,Lindquist}). The FC or dynamic FC serve as sensitive non-invasive measurements to understand disease-related network alterations. Alzheimer’s Disease (AD) is generally regarded as a disconnection syndrome (\citealt{dai2019disrupted}) with gradual network topological changes in a prolonged period with a concealed onset (\citealt{Adelia,Adelib,Romero-Garcia}). Many recent efforts have been put forth to understand the neural underpinning of AD in its early, preclinical stage, also known as mild cognitive impairment (MCI) (\citealt{Misra,Gauthier,Petersen,schwab2018functional,binnewijzend2012resting}), with brain network modeling using graph theory (\citealt{hojjati2017predicting}).

Most complex brain network studies on the AD-related alterations have been largely based on characteristic path length (i.e., the shortest paths) and its derivatives (e.g., assortativity and resilience) (\citealt{ newman2006modularity, ravasz2003hierarchical, achard2006resilient, newman2002assortative, kasthurirathna2013influence}). For example, the averaged characteristic path length of all pairs of brain regions characterizes network’s global efficiency, while the shortest path-based local connectedness defines local clustering coefficients or local efficiency. Recent studies also broadly defines small-worldness, an important brain network property balancing local integration and global reachability (\citealt{Stam}). 
AD has been usually associated with disconnected or less efficient connectome with suboptimal organization (\citealt{Prasad,Supekar,Zippo,Dennis}). MCI, on the other hand, usually manifests increasing FC and suboptimal small-worldness (\citealt{yao2010abnormal,yao2018learning,zhou2011small}), possibly due to a compensatory effect for maintaining normative cognition. While these studies jointly indicated that {\it cost efficiency} can be a good property of the brain network where brain regions are optimally connected to work efficiently together, this phenomenon can be also prominent when only  network’s {\it backbone} (e.g., the top 5\% strongest connections) considered (\citealt{ma2018enhanced,latora2001efficient}). The brain network’s topology can manifest different properties when viewed with more redundant (but weaker) links, which could be equally important as the efficiency-based metrics in understanding the neural mechanisms of diseases (\citealt{bullmore2009complex, wang2015gretna, wangcorrect}).

\begin{figure}[ht]
\centering
\includegraphics[scale=.4]{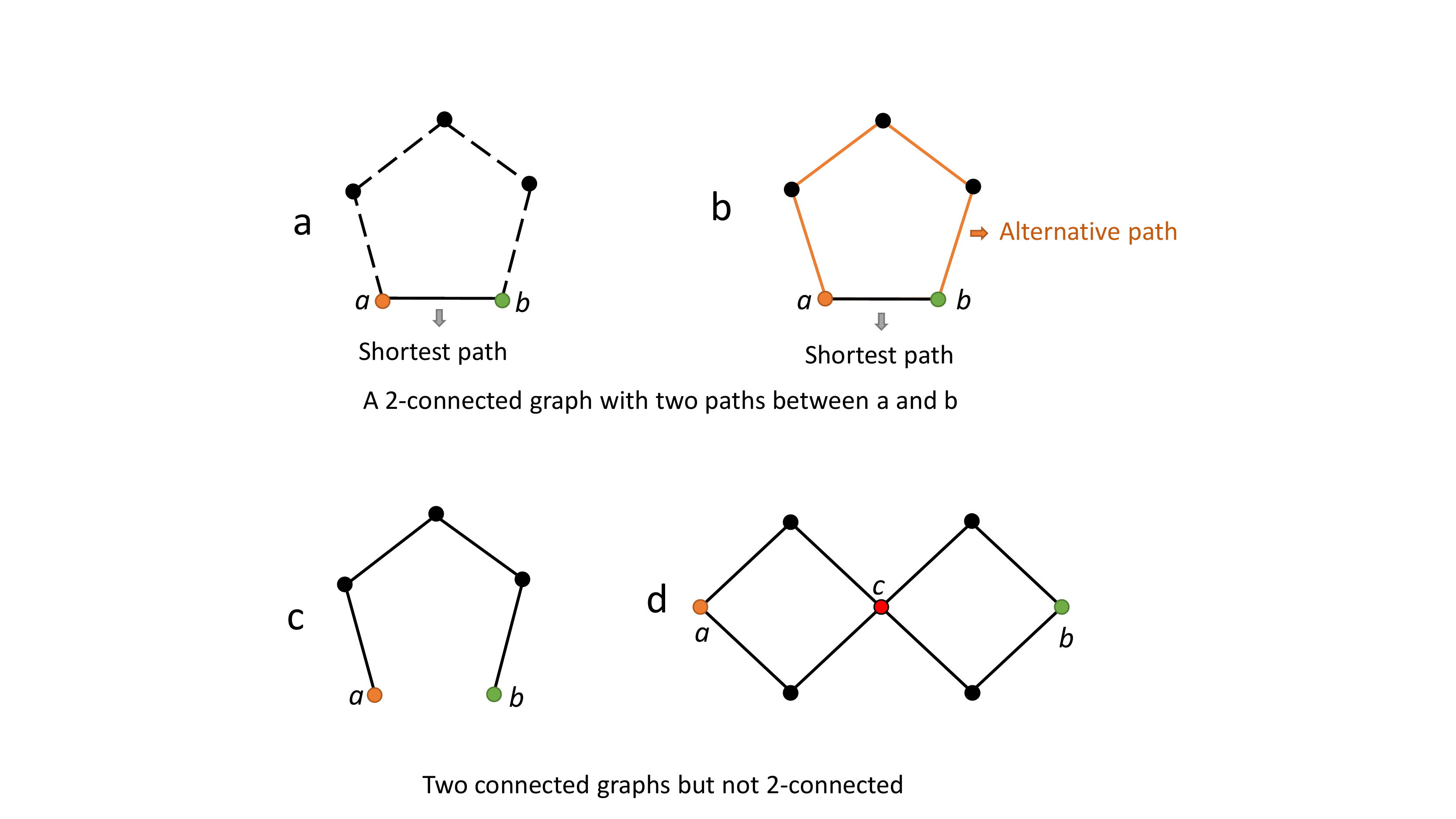}
\caption{Examples of different connected and 2-connected networks (see main text for details)
} 
\label{con}
\end{figure}

In this paper, we investigated brain network changes during AD progression by using new brain network topological metrics that are different from the conventional cost-efficiency methods. We did not only rely on the shortest paths (Fig. \ref{con}a) but also the less investigated (\citealt{di2012redundancy}), alternative (or parallel, see a toy example in Fig. \ref{con}b) paths between each pair of nodes. For example, a magnetoencephalography (MEG)-based FC study defined various redundancy metrics at each frequency band and found that the brain functional network express more redundancy than the random network (\citealt{di2012redundancy}). They further calculated an average number of alternative paths in an electroencephalogram (EEG)-based FC network in the entire network of spinal cord injured patients as a global measurement of redundancy but did not find any significant changes compared to the healthy subjects (\citealt{fallani2011multiple}). In an AD vs. control study, education level was found to act as a cognitive reserve by strengthening the redundancy of a partial brain network constructed by using diffusion MRI (\citealt{yoo2015network}). The inclusion of redundant paths can imply different aspects to the brain connectome, which is further summarized as different types of efficiency, together describing trade-offs among efficiency, cost, and resilience (\citealt{avena2018communication}). Different from these previous studies, we proposed an intuitive and easy-to-calculate metric describing whether  there are {\it always} other paths in addition to the shortest path for {\it every pair} of nodes. Such a redundancy design in other natural networks has been consistently found and studied (\citealt{corson2010fluctuations,harkegaard2005resolving,steiglitz1969design}). We hereby analyzed much denser brain networks by also including weak edges, where the backup routes will likely emerge and the redundancy could become the dominant theme compared to cost efficiency. 

Theoretically, a network $G$ with an edge set $E(G)$ and a node set $V(G)$ is {\it connected} if there exists a path for any two nodes $a, b \in V (G)$. A connected network $G$ becomes {\it 2-connected} if   for every node $x \in V (G)$, $G \setminus x$ is connected ($\setminus$ denotes the removal of a node and all the edges adjacent to it). A 2-connected network has better robustness with a merit of the redundancy design (a cycle consisting of any two nodes) compared to the connected network no matter where the attack occurs. By increasing network density, brain network will change from disconnected to connected and to 2-connected. Thus, two critical points may exist during such a transition: the density that a network first becomes connected from disconnected and the density when it transits from connected to 2-connected. 2-connected network has a good property that, if any shortcut is removed, all the regions are still connected (Fig. \ref{con}a,b). Of note, if multiple paths exist between $a$ and $b$ but all of them share the same node $c$ (Fig. \ref{con}d), they are not redundant paths and the network is not 2-connected. Again, this is not a resilient network because any attack on the node c will break the network. From Fig. \ref{con}, it is easy to know that the network (2-)connectedness can provide a sensitive measurement of AD-related changes as a tiny rewiring to the network could dramatically change its (2-)connectedness.

Since AD progression is a spectrum with gradual changes starting from NC converting to MCI, characterizing (2-)connectedness of the brain FC networks at different stages could better help to understand how AD progression impacts the brain. Due to the findings that the brain FC network changes its topology to meet the moment-to-moment requirement for adaptive thinking and other high-order cognitive functions (e.g., attention and alertness) (\citealt{6,7,5,8}), we further assumed that the brain could be (2-)connected at different density levels in different period of time. Thus, we analyzed (2-)connectedness and their changes in brain {\it dynamic} FC networks. Specifically, we separately assessed connectedness and 2-connectedness of the time-varying FC networks with varied network densities to reveal the aforementioned critical points in a time-resolved manner in NC, MCI, and AD, with specific focuses on {\it both} the transitions from disconnected to connected states and those from connected to 2-connected states, and vice versa. Since AD is considered as a disconnected syndrome, we hypothesized that the connectedness status has been altered in AD versus NC and that the brain chronnectome in MCI might generally have more frequently increased redundancy as a compensatory effect to maintain normative cognitive abilities.

\section{Methods}

\subsection{Data}
In this study, we apply our method to the Alzheimer’s Disease Neuroimaging Initiative (ADNI) datas (http://adni.loni.usc.edu/). Launched in 2003, the original goal of ADNI was to define imaging biomarkers for use in clinical trials of AD. The current goal has been extended to discover more effective methods to detect AD earlier at its pre-dementia stage.  Data quality control was carefully conducted in the ADNI projects to make sure all the data from different imaging centers have the same imaging quality (\citealt{jack2008alzheimer}) (e.g., same imaging protocol, same scanner, and comparable signal-to-noise ratio). The 7-min rs-fMRI data (140 volumes) was preprocessed using AFNI (\citealt{cox1996afni}) according to a standard pipeline (\citealt{yan2010dparsf}). Specifically, the first ten volumes are discarded, followed by a rigid-body head motion correction and a nonlinear spatial registration to the Montreal Neurological Institutes (MNI) space. Frame-wise displacement (FD $> 0.5$) was considered as excessive head motion and the subjects with more than 2.5-min data (50 volumes) labeled as excessive head motion were discarded (\citealt{power2014methods}). 
 FC was assessed using a functional brain atlas (\citealt{shen2013groupwise}) consisting of 268 nodes covering the entire brain. Mean rs-fMRI time series of each brain region was band-pass filtered (0.015–0.15 Hz) and further processed to reduce artifacts by regression analysis (nuisance regressors include head motion parameters (the “Friston-24” model), the mean BOLD signal of the white matter, and that of the cerebrospinal fluid).

 In the first run of analysis, we compared the dynamic properties of connectedness and 2-connectedness between NC, MCI and AD subjects as a main study to understand how dynamic brain functional network changes its redundancy during AD progression. The subjects were selected from ADNI-Go and ADNI-2 only including the baseline scans and ensuring age ($p$ = 0.752, one-way Analysis of Variance (ANOVA), Table~\ref{details}) and gender matched among all three groups. Due to the limited sample size of ADs, we selected the same amount of NC and MCI subjects to make sure as many matched data as possible were used. 

\begin{table}[ht]
   \centering
   \caption{Demographic characteristics (Mean $\pm$ SD) of NC, MCI, AD (used in the main analysis) as well as EMCI and LMCI subjects (also included in the validation analysis) }
  \begin{tabular}{llllllllllll|}
   \hline\noalign{\smallskip}
    & Gender (Male/Female)  & age (year)   &MMSE \\
 \noalign{\smallskip}\hline\noalign{\smallskip}
    \rm NC
&\hspace{6mm}49 (26M, 23F)  &73.1$\pm$6.5    &29.1$\pm$0.9 \\
         \rm MCI
&\hspace{6mm}49 (26M, 23F) &74.3$\pm$9.8     &27.9$\pm$1.6\\
           
   \rm AD
&\hspace{6mm}49 (26M, 23F) &73.3$\pm$8.5      &23.1$\pm$2.5\\
\rm EMCI
&\hspace{6mm}49 (26M, 23F) &74.1$\pm$7.6  &28.2$\pm$1.8\\

\rm LMCI
&\hspace{6mm}49 (26M, 23F) &72.8$\pm$7.7    &26.1$\pm$1.9\\
   \noalign{\smallskip}\hline
       \end{tabular}
\label{details}
\end{table}

\subsection{Overview of the dynamic (2-)connectedness analysis }
\begin{figure}[ht]
\centering
\includegraphics[scale=.35]{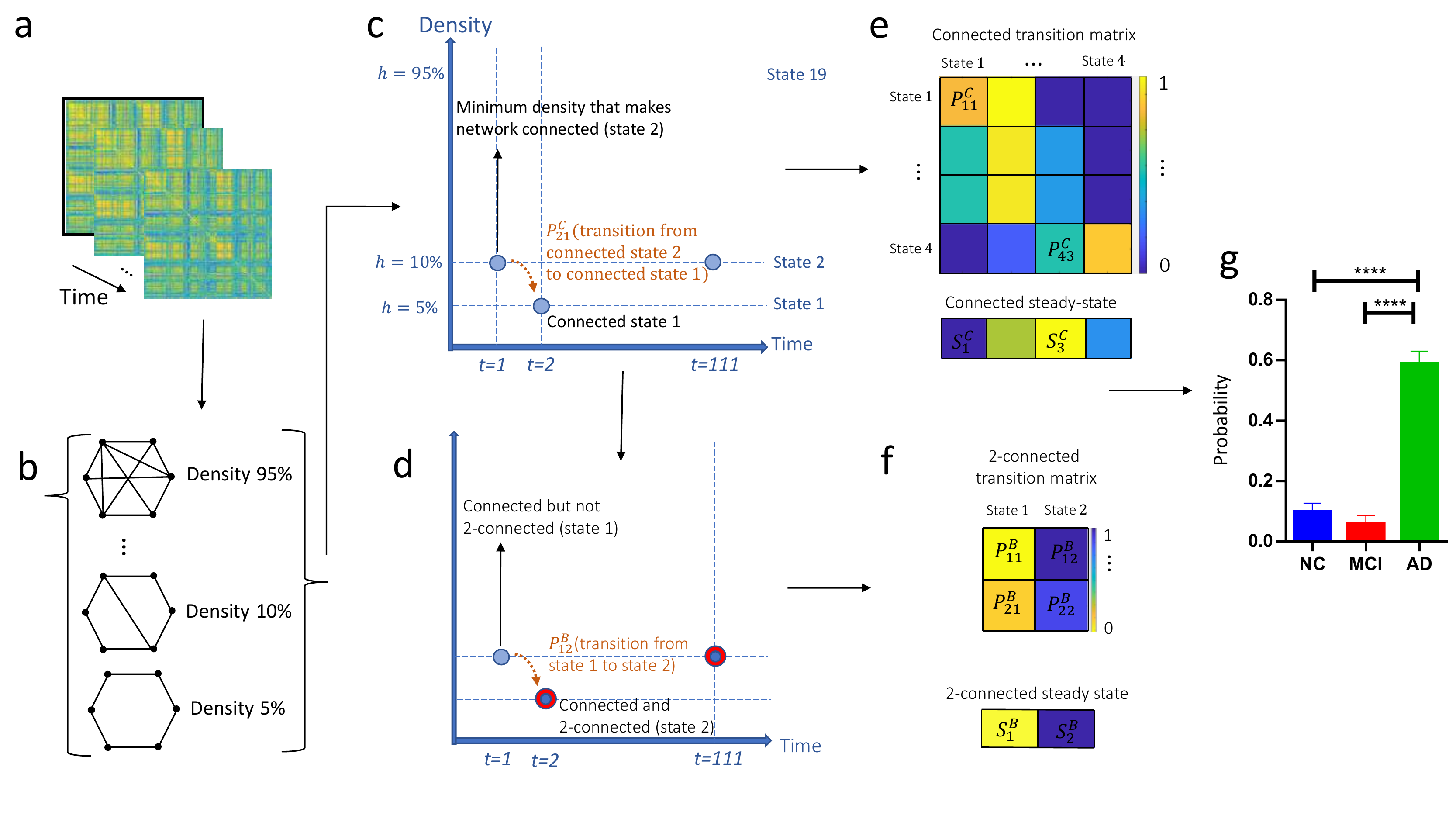}
\caption{ \small The framework of the dynamic (2-)connectedness analysis. (a) Sliding window correlation-based dynamic FC analysis; (b) Constructing binary networks with different density levels for each sliding window-derived network; (c) Calculating a state vector for time-varying connectedness, where blue points represent connected networks at the minimum density level for every time window; (d) Calculating another state vector for time-varying 2-connectedness by re-visiting the previously detected connected states to check whether they are 2-connected at the same time; (e) and (f) Calculating the transition matrix from the connected and 2-connected state vectors; and (g) Conducting group comparison analysis for each time-varying (2-)connectedness feature among NC, MCI and AD groups  } 
\label{brain}
\end{figure}

All the analyses were implemented in MATLAB 2017b, SAGE 8.6, Python 2.7, and SPSS 23. Our method includes three parts: 1) calculating dynamic FC networks; 2) assessing connectedness and 2-connectedness at each time window with varied density levels; and 3) comparing dynamic (2-)connectedness among different groups. The flowchart is shown in Fig.~\ref{brain}.

\subsection{Dynamic FC network construction}
For each subject, a sliding window strategy was used to calculate dynamic FC networks with a window length of 20 volumes (60 s) and a step size of 1 volume (3 s) (\citealt{leonardi}). For the BOLD rs-fMRI signals within each window, pairwise Pearson’s correlation was used to calculate brain FC between every pair of the 268 nodes (Fig. \ref{brain}a). For each of the time-varying FC networks, we calculated its (2-)connected properties at each density level, resulting in a (2-)connectedness property time series for each subject. By definition, connectedness and 2-connectedness are derived from binary networks, where network density is an important parameter that will affect such properties. With a lower density (fewer edges), a network is less likely to be (2-)connected, and vice versa. Searching for a critical point of network density around which the network changes its (2-)connectedness are essential for sensitive group comparisons and for avoiding network saturation at both ends of densities. We applied $H$ different density thresholds to each weighted dynamic FC network to generate $H$ binary networks in each of the $T$ sliding windows (Fig. \ref{brain}b). In this study, $H$ = 19 (from 5\% to 95\% with a step size of 5\%) and $T$ = 111 (windows). 
\subsection{Characterizing (2-)connectedness states }
Denote these networks by $G_{th}$, $1 \le t \le T$ and $1 \le  h \le H$. In each window, let $h_c$ be the minimum density that makes the corresponding network connected. With $h > h_c$, the networks will be all connected with more redundancy;  with $h < h_c$, the networks are all disconnected. The connectedness and 2-connectedness were assessed at such critical points to avoid trivial results and sensitively detect disease-related alterations. Such critical points can be further regarded as “brain states” in terms of the connectedness (Fig. \ref{brain}c) and 2-connectedness (Fig. \ref{brain}d) at a certain period. Denote such a critical network by  $G^*_t$. Like the previous dynamic FC analysis that identifies certain “brain state” for a time window t, we defined a  {\it connected state} for $G^*_t$ and assign $h_c$ ($1 \le h_c \le H$) as its {\it state label}. For example, as shown in Fig. \ref{brain}c, the network’s connectedness in  $t = 1$ is in state \#2, or $C_1 = 2$. We can created a {\it connected state vector}  $\textbf{{C}} := \{C_t \}$ ($1  \le  t  \le  T$) when concatenating the connected states in all time windows, indicating how the dynamic brain network changes its connectedness property. The definition of connected states inherently codes network topology across multiple density settings. 

 We further checked the $G^*_t$ to see if it was also 2-connected. If so, we define the network as it is in a 2-connected state \#2; if not, state \#1. Likewise, we further created a {\it 2-connected state vector} $\textbf{{B}} := \{B_t \}$ ($1  \le  t  \le  T$). For example, $B_1$ = 1 (connected but not 2-connected) and $B_2$ = 2 (both connected and 2-connected) (Fig. \ref{brain}d). The state vector \textbf{{B}} describes how the time-varying brain FC network changes its 2-connected state along time. 
 
 There are 19 possible states (since $H$ = 19) for connectedness. However, we found that most transitions among the connected states occurred between \#1 to \#4; therefore, only the first four connected states were considered in the following analysis. Both of the 2-connected states were used. As the state of 2-connectedness was determined according to the critical point in terms of connectedness, it is irrelevant to a specific network density. Of note, one can define 2-connected states in the same way as that of the connected states; however, since connectedness is a necessary condition for a network to become 2-connected, further checking if the same critical points generates 2-connected network could reveal more sensitive information to subtle changes in the network topology. 
\subsection{Transition among different (2-)connectedness states}
We further quantified the dynamic properties of the connectedness and 2-connectedness states with a transition matrix (describing the probability of one state transitioning to another) and a steady state vector (describing the probability of a certain state be cumulatively occupied given a long enough time, which equals to the normalized ``dwelling time'') based on Markov Chain (\citealt{williams2018markov,chavez2010complex}). This resulted in a  $4 \times 4$ {\it connectedness transition matrix} $\textbf{{P}}^C:=\{ P^C_{ij}\}$ (1 $\le i, j \le$ 4), where $ P^C_{ij}$  indicates the probability of changing from state $i$ to $j$, and a $2 \times 2$, {\it 2-connectedness transition matrix} 
$\textbf{{P}}^B:=\{ P^B_{ij}\}$ (1 $\le i, j \le$ 2). 
The steady state vector is a probability vector $\textbf{{S}}$ that satisfies the equation 
$\textbf{{S.P}} = \textbf{{S}}$, where $\textbf{{P}}$ is the transition matrix. It was solved as the left eigenvector of  $\textbf{{P}}$ corresponding to the eigenvalue of 1. $\textbf{{S}}^C$ and $\textbf{{S}}^B$ have a length of four and two, respectively. Therefore, we generated 26 different features, including 16  $ P^C_{ij}$ in $\textbf{{P}}^C$ and $\{S^C_1,S^C_2,S^C_3,S^C_4, P^B_{11},P^B_{12},P^B_{21},P^B_{22},S^B_{1},S^B_{2}\}$ for every subject. Please note that $P^B_{11}$ and $P^B_{12}$, $P^B_{21}$ and $P^B_{22}$, as well as
$S^B_{1}$ and $S^B_{2}$ indicate the same features, because 
$P^B_{11}+P^B_{12}=1$, $P^B_{21}+P^B_{22}=1$, and $S^B_{1}+S^B_{2}=1$. Therefore, only 23 independent features needed to be considered, containing 16  $ P^C_{ij}$ in $\textbf{{P}}^C$ and $\{S^C_1,S^C_2,S^C_3,S^C_4, P^B_{11},P^B_{12},S^B_{1}\}$. However, for completeness, we considered all the 26 features in tables and figures in the following analysis.

\subsection{Statistical comparisons among NC, MCI, and AD  }
For each of the 26 dynamic connectedness features, we conducted a Kruskal-Wallis test (a non-parametric version of the one-way ANOVA) to detect group differences among NC, MCI, and AD groups. Family Wise Error (FWE) corrected $p <$ 0.05 was used to indicate significant group differences. Mann-Withney U-tests (a non-parametric version of the two-sample t-test, two tailed) were further used to conduct {\it post hoc} pairwise comparisons for the significant Kruskal-Wallis test results ($p <$ 0.05, FWE corrected).

 \subsection{Validation with two independent MCI subgroups  }
 To further validate the main results and to see if the revealed abnormalities can be detected at an even earlier stage of MCI, we conducted a second analysis by replacing the MCI group with an independent dataset consisting of two different age- and gender-matched MCI subgroups (early(E-) and late(L-) MCI  (\citealt{edmonds2019early}) selected from the ADNI GO/2 (Table \ref{details}). Therefore, we had four different groups (NC, EMCI, LMCI, and AD) that allowed us  to further reveal the gradual changes and {\it earlier} signs (as shown in the EMCI group) of the AD progression. The NC and AD groups are the same as those in the main analysis. The four groups have comparable age ($p$ = 0.872, one-way ANOVA) and with gender matched. All the data analyses are kept the same as those in the main analysis. The results were compared with the main ones for validation, with a particular focus on whether there were significant differences between NC and EMCI and whether the general results still held from NC to AD.

 \subsection{Test-retest reliability assessment}
 We conducted a third analysis by evaluating test-retest reliability of our method and to check if the main findings could be replicated using another follow-up scan from the same subject. Specifically, we identified a follow-up dataset from several of the NC, EMCI, LMCI and AD subjects in Table \ref{details}. A total of 14 subjects (7 males and 7 females) from each group have available retest scans and the average test-retest interval for NC, EMCI, LMCI and AD groups are 9.9$\pm$7.6, 7.4$\pm$3.9, 8.8$\pm$3.9 and 4.8$\pm$1.4 months, respectively. Retest samples are age  (p = 0.329, one-way ANOVA, two-tailed with degree of freedom 3) and gender matched.

\subsection{External validation with independent dataset}
In addition to the ADNI dataset, we also used an independent dataset consisting of 67 NC (65.9$\pm$7.2 years old, M/F: 31/36, MMSE = 28$\pm$2.1) and 71 amnestic MCI subjects (68.3$\pm$9.4 years old, M/F: 33/38, MMSE = 24.4$\pm$3.4) and from Xuanwu hospital of Capital Medical University as an out-of-sample validation and reproducibility assessment of our method. The diagnosis of amnestic MCI patients was met the criteria proposed by (\citealt{petersen2001current,petersen2004mild}). The details of the data (e.g., inclusion/exclusion criteria, rs-fMRI protocols) were described (\citealt{chen2016neuroimaging}).
 These two groups have comparable age ($p$ = 0.1, 2-sample ttest) and gender ($p$ = 1, chi-square test). This dataset is not included in any centers of ADNI project. The same analysis was conducted comparing the differences in 2-connected steady states ($S_1^B$ and $S_2^B$, as the main analysis indicated that MCI tended to have altered 2-connected steady states compared to NCs).

\section{Results}
\subsection{Dynamic brain networks in AD are more likely connected in sparse settings}
\begin{figure}[ht]
\centering
\includegraphics[scale=0.36]{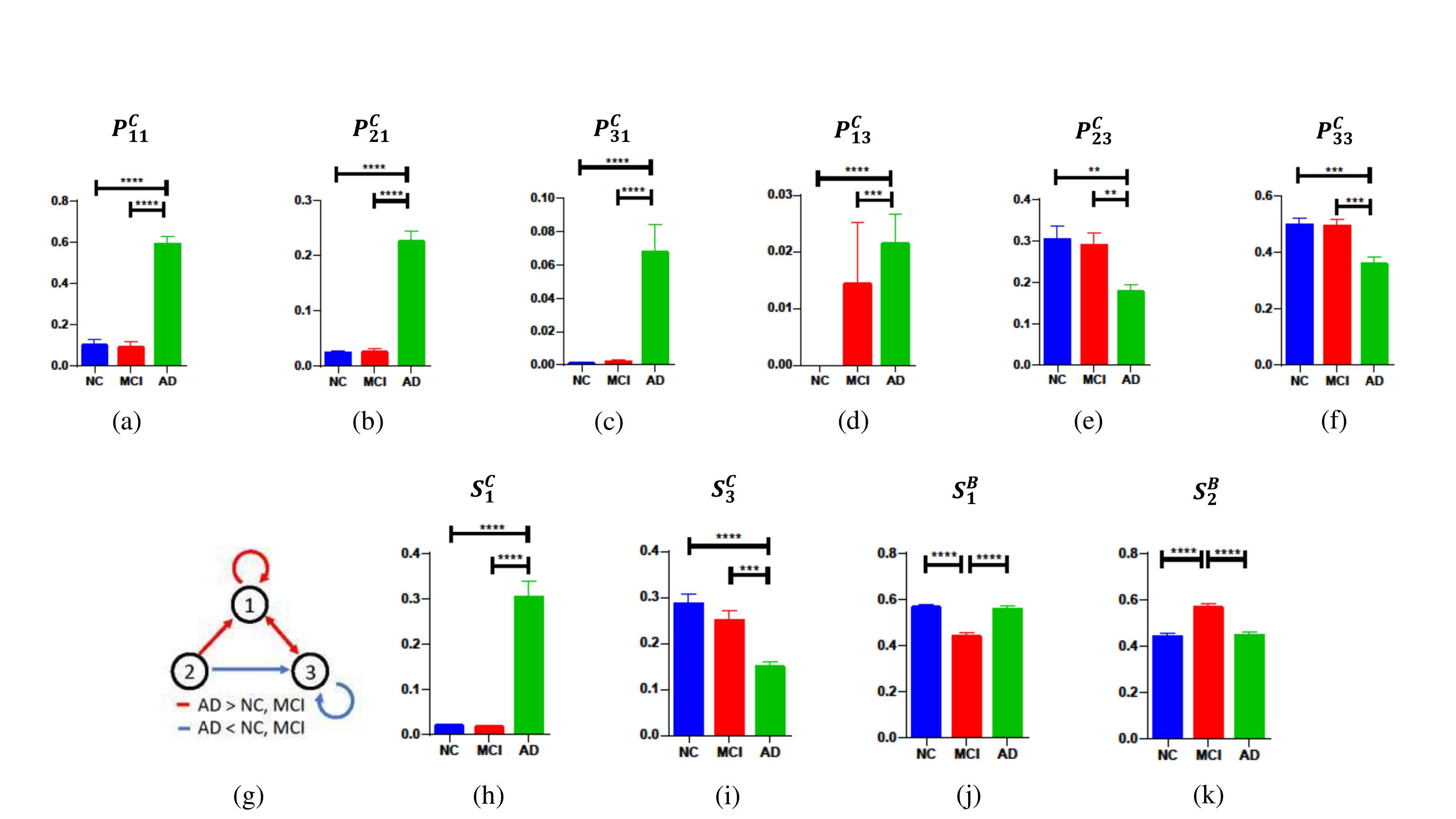}
\caption{ \small Group comparisons of each dynamic (2-)connected metrics among NC, MCI, and AD groups, where *, **, ***, **** indicate corrected $p$-values at the intervals of (0.01,0.05], (0.001,0.01], (0.0001,0.001], and (0.00001,0.0001], respectively; error bars show standard errors (SE). Subplot (g) summarize major differences between AD and NC/MCI in terms of the transition probabilities of the connectedness states \#1-3 (a-f) as well as the steady connected states (h-i) } 
\label{Fig1}
\end{figure}

As shown in Table \ref{test3}, 10 out of the 26 features had significant group differences as detected by the three-group comparisons based on Kruskal Wallis tests. Post-hoc analysis revealed that the group differences were mainly contributed by AD (i.e., differences were mainly found in AD vs. NC and in AD vs. MCI), except the 2-connected steady states  $ S^B_{1}$ and $S^B_{2}$, in which MCI also showed differences from NC (Table \ref{test2}, Fig.  \ref{Fig1}). 
\begin{table}[ht]
   \centering
    \caption{Results from Kruskal-Wallis tests comparing dynamic (2-)connectedness among NC, MCI and AD groups (bold numbers denote significant results after family-wise error corrections)  }
  \begin{tabular}{llllllllllll|}
    \hline\noalign{\smallskip}
    & $P^C_{11}$  & $P^C_{21}$  & $P^C_{31}$  & $P^C_{41}$  & $P^C_{12}$  & $P^C_{22}$  & $P^C_{32}$  & $P^C_{42}$  & $P^C_{13}$  & $P^C_{23}$   \\
   \noalign{\smallskip}\hline\noalign{\smallskip}
    \rm NC/MCI/AD
&{\bf.000}   &{\bf.000} &{\bf.000}  &.368  &.167 &.038  &{.016}   &.578  &{\bf.000}  &{\bf.001}  \\
      \noalign{\smallskip}\hline
       \end{tabular}
       
       \begin{tabular}{llllllllllll|}
   & $P^C_{33}$  & $P^C_{43}$  & $P^C_{14}$   & $P^C_{24}$  & $P^C_{34}$  & $P^C_{44}$  & \hspace{1mm}$S^C_1$  & \hspace{1mm}$S^C_2$  & \hspace{1mm}$S^C_3$  & \hspace{1mm}$S^C_4$   \\
      \noalign{\smallskip}\hline\noalign{\smallskip}
    \rm NC/MCI/AD
&{\bf.000}	&.293 &.048 &.552 &.861 &.173 	&{\bf.000}	&.875	&{\bf.000}	&.017 \\
     \noalign{\smallskip}\hline
       \end{tabular}
        
         \begin{tabular}{llllllllllll|}
    & $P^B_{11}$ & $P^B_{12}$ & $P^B_{21}$ & $P^B_{22}$  &\hspace{1mm} $S^B_{1}$ &\hspace{1mm} $S^B_{2}$
\\
   \noalign{\smallskip}\hline\noalign{\smallskip}
   \rm NC/MCI/AD
&.892	&.892	&.991	&.991	&{\bf.000}	&{\bf.000} \\ 
    \noalign{\smallskip}\hline
       \end{tabular}
 
\label{test3}
\end{table}
\begin{table}[ht]
   \centering
   \caption{Pairwise comparison results with Mann-Withney U tests on the significant dynamic (2-)connectedness features from Table \ref{test3} (bold numbers denote significant results after family-wise error corrections) }
  \begin{tabular}{llllllllllll}
     \hline\noalign{\smallskip}
     & $P^C_{11}$  & $P^C_{21}$  & $P^C_{31}$ & $P^C_{13}$ & $P^C_{23}$   & $P^C_{33}$    & \hspace{1mm}$S^C_1$  & \hspace{1mm}$S^C_3$
     & \hspace{1mm}$S^B_{1}$ &\hspace{1mm} $S^B_{2}$
     \\
   \noalign{\smallskip}\hline\noalign{\smallskip}
    \rm AD/NC &{\bf.000}  &{\bf.000}	 &{\bf.000}		 &{\bf.000}		 &{\bf.001}	  &{\bf.000}		&{\bf.000}	&{\bf.000} 	&.815	&.815 \\
    \rm  AD/MCI &{\bf.000}	&{\bf.000}	 &{\bf.000} 		 &{\bf.000}		 &{\bf.002}	&{\bf.000}		&{\bf.000}		&{\bf.000} &{\bf.000}	&{\bf.000} \\ 
    \rm  MCI/NC &.661 &.997 &.310  &.155 &.619  &.972 &.916 &.251  &{\bf.000} &{\bf.000} \\
      \noalign{\smallskip}\hline
       \end{tabular}
        
\label{test2}
\end{table}
For transitions among the connected states, we found that, as long as the connected state \#1 (i.e., the brain network is connected at a very sparse setting with only 5\% edges) was involved, AD tends to have larger transition probabilities    ($P^C_{11},P^C_{21},P^C_{31}$ and $P^C_{13}$  , Fig. \ref{Fig1}a-d)
 compared to NC and MCI. These results further led to a more steady connected state \#1 in AD  ( $S^C_1$, Fig. \ref{Fig1}h). These results indicate that AD tends to spend a longer time with the connected network even under a very sparse setting compared to the other groups. Of note, such results does not mean AD tends to have stronger brain dFC; instead, they altogether mean that the AD’s brain tends to have altered dFC that re-distributed the strongest edges to form a more evenly pattern that made the dynamic network more likely connected. On the other hand, AD subjects have to make big changes in terms of the network density (e.g., change from 5\% to 15\%, and vice versa, Fig.  \ref{Fig1}c, d) to maintain connectedness status.

If the connected state \#3 (i.e., the brain network is connected at a not quite sparse setting with 15\% edges) was involved and the previous connected states are also not with a quite sparse setting (10\% and 15\%, rather than 5\%), AD tends to have lower  $P^C_{23}$ and $P^C_{33}$, (Fig. \ref{Fig1}e, f), possibly due to the dominant findings for $P^C_{13}$ and $P^C_{31}$. All the results from the connected state transition analysis are summarized in the schematic plot (Fig. \ref{Fig1}g).

Different from the fact that AD’s brain network tends to be connected at a low redundancy situation ($h$ = 5\%), NC and MCI tend to have connected brain networks in higher redundancy scenarios ($h$ = 15\%, Fig. \ref{Fig1}i). This indicates that NC and MCI (especially NC) tend to spend more time with the connected network under more redundant network settings compared to AD (possibly due to the strong FCs are more distributed within such functional sub-network, which makes the entire network less likely to be connected in a sparse setting). 
\subsection{2-connected states could be used to differentiate MCI from NC }
When we checked if the network was, at the same time, 2-connected when it first became connected at the critical point, MCI’s brain network showed a more likely tendency to be also 2-connected compared to NC and AD, with the latter two groups showing similar results (Fig. \ref{Fig1}k). In other words, at the critical point, MCI’s brain network was less probable to be only connected but not 2-connected compared to NC and AD (Fig. \ref{Fig1}j). Of note, both Fig. \ref{Fig1}j and Fig. \ref{Fig1}k describe the same difference. However, none of the 2-connected state transition $P^B_{ij}$ showed any group difference. As the steady 2-connected state showed a significant group difference between NC and MCI, this feature might be adopted to detect AD at its preclinical stage.  

\subsection{Validation analysis}
We successfully validated the main results by replicating the analysis on newly included subdivided MCI groups (EMCI and LMCI). With MCI replaced with two subgroups, the trends among NC, (E/L)MCI, and AD were largely similar (see Online Appendices) compared to the main results in Tables \ref{test3} and \ref{test2}, except  $P^C_{22}$ and $S^C_4$ also showed differences among the four groups. The results of EMCI and LMCI were very similar to each other and all together similar (see Online Appendices) to the results of the single MCI group (Fig. \ref{Fig1}). Collectively, separating the MCI into EMCI and LMCI did not change the main conclusions.

More interestingly, we spotted a trend (although it was not significant) with continuous changes from EMCI to LMCI and then to AD (see Online Appendices), especially those for  $P^C_{13}$ and $S^C_3$ (see Online Appendices). They indicate that there could be gradual changes in terms of the connected state transitions and LMCI subjects are closer to AD than EMCIs. More importantly, as shown in Online Resource the differences in the steady states of 2-connectedness ($S^B_1$ and $S^B_2$) between MCI and NC were preserved in the result of EMCI vs. NC, indicating that the detection of potential AD might be achieved at an even earlier (EMCI) stage.

\subsection{Test-retest reliability}

\begin{figure}[ht]
\centering
\includegraphics[scale=0.35]{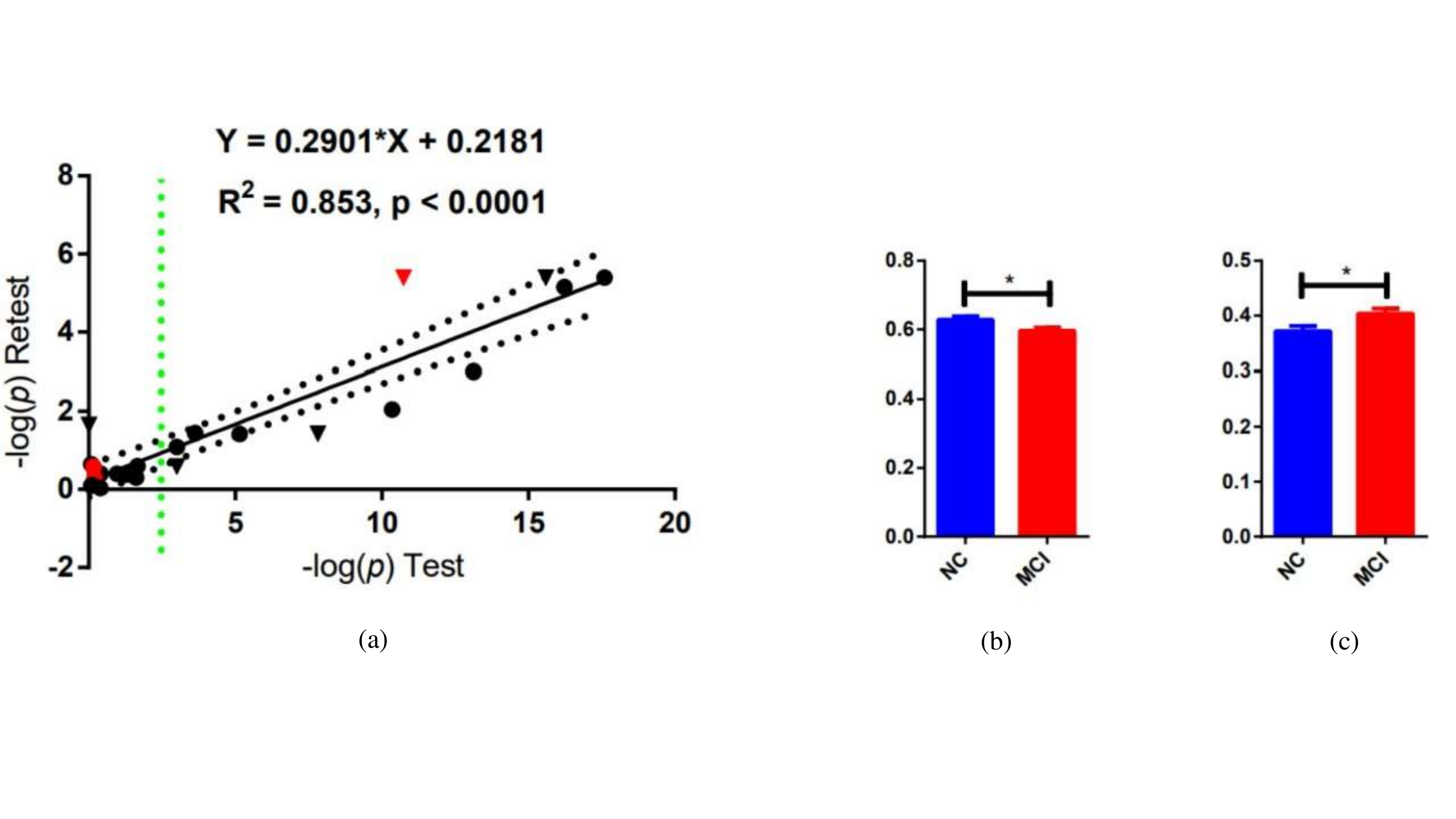}
\caption{ \small (a) Scatter plot of the four-group comparison results of the 23 (after removing three same dynamic 2-connectedness features, see definitions of 2-connectedness features) dynamic (2-)connectedness features based on the test and retest data where $-log(p)$ quantifies the strength of the group differences. Black and red colors show connected and 2-connected features and round and diamond shapes indicate transition and steady state features, respectively. The green dashed line represents the threshold $p$ = 0.05/23 applied to the test data. (b-c) Comparisons of dynamic 2-connected features, $S^B_1$ and $S^B_2$, between NC and amnestic MCI subjects from an independent imaging center (Xuanwu Hospital) for external validation, where * indicates $p <$ 0.05
} 
\label{log}
\end{figure}

After repeating the same analysis on the follow-up data from a subset of each of the four groups used in the validation analysis, we found largely similar results (see Online Appendices). As shown in Online Resource, most of the (2-)connectedness features from retest data are similar to the results from the test data. We further plotted the strength of the group differences (as defined by transforming the Kruskal Wallis test-derived $p$-values using $-log(p)$, with a smaller the $p$ indicating a larger group difference) from the test data against those from the retest data. The result indicated an excellent test-retest reliability (with a Spearman correlation $r$ of 0.922) of our method in detection AD-related dynamic network redundancy differences  (Fig. \ref{log}).

\subsection{External validation with independent dataset}

From the main results, we found that NC and MCI showed significant differences in steady 2-connected states $S^B_1$ and $S^B_2$. Then we checked these two features on the independent data consisting of NCs and amnestic MCIs from Xuanwu Hospital as an external validation. Since  $S^B_2$ is essentially reflecting the same difference as $S^B_2$ , only one feature was considered. Therefore, no multiple comparison correction was needed. We found that $S^B_1$ (and $S^B_2$) showed a significant difference ($p$ = 0.015, Fig. \ref{log}b, c) between NCs and amnestic MCIs, similar to the previous findings between NCs and MCIs based on ADNI datasets (Fig. \ref{Fig1}j, k). 

 \section{Discussion}
 
 In this paper, we adopted novel brain network attributes, connectedness and 2-connectedness, to quantify dynamic changes in brain chronnectome among different groups at different stages of AD progression. We found that these measurements were able to sensitively detect AD-related topological changes in the spatiotemporal brain functional network patterns. Our major findings are as follows. {\it First}, we found that AD subjects had a more frequently connected brain network in the sparse setting (Fig. \ref{Fig1}a-c, h). It indicates that AD patients tend to maintain a connected brain network for longer time compared to NC and MCI when only the network backbone is considered. {\it Second}, we found that AD subjects manifested large temporal fluctuations in terms of the critical points (e.g., with the minimal network density changing from 5\% to 15\% to maintain connectedness in consecutive temporal windows, Fig. \ref{Fig1}c, d) more frequently than NC and MCI. {\it Third}, by considering 2-connectedness, subjects with MCI tend to be also 2-connected more frequently at the critical point of connectedness (Fig. \ref{Fig1}j, k). The findings imply that, while the MCI’s brain network (together with the NC’s) tends to become connected with denser edges compared to AD’s, it becomes too complete such that the entire network is more likely to be redundant than AD’s network.
 
The results jointly indicated gradually altered brain network topology in two folds. On one hand, although AD is generally regarded as a disconnection syndrome (with globally decreased FC compared to NC and MCI), they tend to be more connected in a low density level. This result does not contradict previous findings. This is because the averaged FC weights in AD could be lower than those in NC and MCI, but the strongest FC links in AD could be distributed more evenly from intra- to inter-sub-network connections, making the entire network more likely to be connected. Such a status may not be optimal and stable, as the AD’s brain network can sometimes be quite disconnected and needs more weak connections to be included (a higher density) to maintain connectedness (Fig. \ref{Fig1}a-d). On the other hand, when considering 2-connected state with redundant connections, subjects who could be in an early stage of AD (MCI) show more differences compared with NC, where the MCI’s brain network features a more robust and redundant topology (larger probability of steady 2-connected state, Fig. \ref{Fig1}k) compared with NC and AD. Such an increment of redundancy may help to ensure a back-up path for every shortest path and maintain the network’s efficiency even under random AD pathological attacks. As such a phenomenon was only detectable in MCIs, it could be interpreted as an overshoot with protective and compensatory effects for MCIs to maintain their cognitive level in the presence of AD-related neurodegeneration (\citealt{ petersen2014mild}). More importantly, the MCI’s abnormally elevated redundancy can be detected at an even earlier (i.e., early MCI) stage, indicating good sensitivity of the proposed 2-connectedness measurements.  

It is noticeable that the major findings from the multiple ADNI centers are test-retest reliable and reproducible, with an acceptable external validity based on an independent dataset. For example, we found a significant difference between MCI and NC in steady 2-connected state $S^B_1$ based on the main test (the probabilities for NCs and MCIs are 0.56$\pm$0.10 and 0.44$\pm$0.11), while that from the validation analysis and the retest dataset are very similar (0.56$\pm$0.10 vs. 0.44$\pm$0.08 for NCs and EMCIs, and 0.58$\pm$0.06 vs. 0.38$\pm$0.11 for the retest data from the same NCs and EMCIs). An independent dataset from Xuanwu Hospital revealed the same trend despite an elevated baseline, where the steady 2-connected state $S^B_1$ from NCs (0.62$\pm$0.08) were still higher than that from amnestic MCIs (0.59$\pm$0.07). The effect size on this difference according to Cohen’s d are large (1.14 for main test, 1.32 for validation analysis, and 2.25 for retest, despite a medium one for the independent test (0.40). The decreased effect size may come from different ethnicities of the subjects from ADNI (mostly Caucasians) and Xuanwu data (all Chinese), different imaging protocols and scanners (e.g., rs-fMRI temporal resolution), different ages and gender ratio (the selected subjects from Xuanwu Hospital are younger than selected ones from ADNI), and/or different MCI diagnostic criteria. Further tests on even larger dataset is necessary to validate our findings.

   Graph theory has become a powerful tool in the investigation of brain topological changes in diseased populations including AD (\citealt{karwowski2019application}). While some previous studies use connectedness as a prerequisite condition to quantify other network attributes (e.g., shortest path length)  (\citealt{ meier2015union}), they did not directly compare the connectedness. None of them has ever focused on more redundant network topology such as 2-connectedness. While a handful of previous works have investigated alternative paths  (\citealt{yoo2015network}), none of them examined more stringent (2-)connected properties where all pairs of regions must have one (or more than one) independent paths. Our results suggested that these stringent global properties could be quite sensitive in tracking AD progression. We showed the first-ever evidence that time-varying brain connectedness can be informative to understand AD progression. While the majority of the complex brain network studies focused on the  shortest pathsregions (\citealt{ meier2015union})  and the derived metrics (e.g., betweenness centrality (\citealt{rubinov2010complex}) and connector hub (\citealt{van2013network}), we showed that other lengthier pathways could be also informative, thanks to the similar concepts and applications in the other fields  (\citealt{corson2010fluctuations,harkegaard2005resolving,steiglitz1969design,white2001fast,quattrociocchi2014self}). To further increase sensitivity, we extend such a concept to dynamic FC network by investigating how the brain changes its {\it reserved paths} in a time-varying manner, making it feasible to detect subtle changes in a diseased condition. The framework can be easily applied to other brain diseases and mental disorders. 
  
  While the current framework focused on dynamic networks, it is straightforward to apply it to static brain network or a structural connectivity network based on diffusion MRI. In addition, 2-connectedness can be further extended to 3-connectedness (at least three independent paths exist for every pair of brain regions) for network studies in a more redundant scenario. The current work characterized the entire brain network; to improve spatial specificity, one can investigate functional sub-networks (e.g., default mode network) or a sub-network associated with each region for a fine-grained investigation and biomarker detection.
  
In conclusion, we used novel network redundancy measurements to reveal how dynamic brain functional network changes its topology in more denser conditions during the AD progression. The reliable and reproducible findings provide a new view angle to the AD-related brain networks and a sensitive means to detect AD in its early stage. We advocate that the redundant design is as important as cost efficiency and could be promising for the future network neuroscience studies.


%
%

\end{document}